\documentclass{article}
\usepackage{spconf,amsmath,graphicx}

\usepackage{enumitem}
\usepackage{microtype}
\usepackage{xcolor}
\setlist{nosep, leftmargin=14pt}

\usepackage{mwe} 

\title{Assessment of deep-learning methods for the enhancement of experimental low dose dental CBCT volumes}
\name{Louise Friot-{}-Giroux, Françoise Peyrin, Voichi\c ta Maxim}
\address{Univ Lyon, INSA-Lyon, Université Claude Bernard Lyon 1, CNRS Inserm,\\ CREATIS UMR 5220, U1294, F-69621, LYON, France}

\begin{document}
%
\maketitle
\begin{abstract}

Cone-beam tomography enables rapid 3D acquisitions, making it a suitable imaging modality for dental imaging. However, as with all X-ray techniques, the main challenge is to reduce the dose while maintaining good image quality. Moreover, dental reconstructions face a series of issues stemming from truncated projections as well as metal and cone beam artifacts. The aim here is to investigate the ability of neural networks to improve the quality of 3D CBCT dental images at low doses. We test different configurations of convolutional neural networks, trained in a supervised way to reduce artifacts and noise present in analytically reconstructed volumes. In a study on 32 experimental cone beam volumes, we show their capacity to preserve and enhance details while still reducing the artifacts. The best results are obtained with a 3D U-Net which compares advantageously with a TV regularized iterative method and is considerably faster. 
\end{abstract}

\section{Introduction}

3D X-ray tomography involves reconstructing an inner volume from its projections. Cone beam CT is increasingly used for various applications and in particular in dental imaging. However, since exposure to X-rays is harmful, one of the major challenges facing X-ray imaging today is to reduce the absorbed dose, while maintaining sufficient image quality for correct diagnosis. 

The most popular reconstruction methods are the so-called analytical methods. In 2D, the filtered back-projection algorithm accurately reconstructs the volume under ideal conditions. However, it is sensitive to noise - and therefore to dose reduction - and is prone to artifacts in geometry-limited acquisitions. A generalization of this algorithm to 3D conical geometry (CBCT) has been proposed by Feldkamp et al \cite{Feldkamp1984} (FDK). In practice, in-patient measurements depend on numerous physical parameters and noise is particularly present in low-dose acquisitions, which limits the performance of these analytical solutions.

Iterative reconstruction methods offer greater flexibility, at the cost of higher computation times. The tomographic reconstruction problem can be modeled by a linear problem, which consists of finding the volume $f$ from the projections $p$, with $A$ the projection operator: $Af = p$. It can be tackled by a variational approach, which aims to solve the inverse problem by minimizing a functional consisting of a data fidelity term and a prior on the desired solution \cite{Chambolle2011, Sidky2012}. Among conventional iterative methods, we can mention SIRT (Simultaneous Iterative Reconstruction Technique) \cite{Gilbert1972} in which the noise on the data is assumed to be Gaussian, MLEM \cite{Lange1984} for Poisson-distributed data, and their respective regularized versions with Total Variation (TV) prior \cite{Sidky2006} and \cite{Maxim2023}.  

The application of deep learning methods to CT has shown superior performance than analytical and iterative methods \cite{Jin2017, Adler2017} in terms of noise suppression from reconstructed images. The main advantage of these methods is that they require neither precise knowledge of the noise present in the data, nor explicit a priori information about the solution. Indeed, in the case of supervised learning, these information are implicitly deduced by the network from the database. Reconstruction time is also an argument in favor of these learning methods compared to iterative methods. 

While deep learning appears to outperform analytical and iterative methods in terms of noise suppression, it is vital to evaluate this performance in clinically relevant conditions, i.e. from real acquisitions.

The aim of this work is to study the ability of a U-Net neural network to remove artifacts and denoise volumes reconstructed from low-dose acquisitions in the case of dental CBCT. Three networks were studied, 2D U-Net, Multi-plane U-Net and 3D U-Net which were trained from our own experimental dental CBCT database. Their performance were evaluated by cross-validation and compared with an iterative reconstruction including a TV (Total Variation) prior. 

\section{Methods}

\subsection{U-Net-2D}

U-Net is a popular CNN that was introduced for segmentation \cite{Ronneberger2015}, but it can also be used to improve image quality. For post-processing the CBCT volumes, we used a slightly modified version of the network proposed by \cite{Jin2017}, a residual U-Net, with a depth of 5, 64 input channels and around 31 million parameters. 

We replaced ReLU activations with Leaky ReLU, which improved training stability. Deconvolution layers, known to create checkerboard artifacts in high-intensity areas \cite{Odena2016}, have been replaced by upsampling layers with nearest-neighbor interpolation. We removed gradient clipping as it created blurred zones around the metal and slowed down convergence. The method was applied slice-by-slice to the axial slices of the FDK reconstructed volume.

\subsection{U-Net-MP}

In the previous 2D network, adjacent slices were processed independently of each other, resulting in defaults that are visible in the cross-section. We then trained three U-Nets, each in a different direction ($\Gamma_A$ in axial, $\Gamma_C$ in coronal and $\Gamma_S$ in sagittal direction), and considered a weighted average of the three results, to which we refer as U-Net-MP (Multi-Planes): $f_{\text{U-Net-MP}} = 0.5 f_{\Gamma_A} + 0.3 f_{\Gamma_C} + 0.2 f_{\Gamma_S}$. This network integrates 3D coherence with a moderate increase in the computational burden. 

The selected weights reflect the differences in performance to take into account the differences in performance between the three networks, related to specificities of CBCT acquisitions. These differences can be explained by the direction of the X-rays during acquisition, which favors the axial direction. In addition, due to the shape of the skull, sagittal slices have a lower SNR, resulting in directional noise. 

\subsection{U-Net-3D and U-Net-3D+}

Finally, a U-Net-3D architecture has been implemented, on the mode of the 2D one, by replacing the convolution and upsampling layers with their 3D counterparts. In order to keep moderate the computing resources demand, and as close as possible to the 2D case, we have divided the number of convolution filters by two, thus obtaining a network of around 21 million parameters. 

The lower performance of this U-Net-3D network compared with $\Gamma_A$ and U-Net-MP led us to also consider a 3D network with the same number of filters as in 2D, i.e. a network with 84 million parameters noted U-Net-3D+ for the comparison results.

\subsection{KL-TV}

For comparison purposes, we considered the KL-TV preconditioned algorithm \cite{Sidky2012} that achieves in \cite{friot2022iterative} competitive results on dental CBCT. This primal-dual algorithm converges towards the minimum of a functional consisting of the Kullback-Leibler divergence for data attachment, and the total variation for regularization:
$$ f^* = \arg\min_f \sum_i \left[ Af - p\log (Af)_i \right] + \alpha \Vert \vert \nabla f\vert\Vert_1 ,$$
where $f$ denotes the volume to reconstruct, $p$ the acquired projections, $A$ the projections operator and $\nabla f$ the gradient of $f$. 
The application of this algorithm in dental CBCT has been detailed in \cite{friot2022iterative}.

\section{Experiments and Results}

\subsection{Database}

We used an experimental CBCT database composed of 32 anonymized patients, of both sexes, aged between 18 and 70 years. Cone beam projections were acquired in normal dose, shortscan by a CS 8100SC 3D scanner. The average volume size was $320\times 280\times 280$~voxels, the size of a voxel being 300 µm$^3$. The FDK reconstruction of these volumes gave us the ground truth, and the low-dose volumes were obtained by analytical reconstruction after sub-sampling the set of projections by keeping each fifth projection. 

The database was divided into twenty-five volumes used for training and a further six for validation and hyper-parameter adjustment. The last volume was used for testing.
Volumes were globally normalized between -1 and 1, independently of the maximum value of each. 

\subsection{Experimental Settings}

The 2D networks were trained over 50 epochs, with a batch size equal to 10. The U-Net 3D were trained over 100 epochs, with a patch of size $160\times 160\times 160$ for each batch for the U-Net-3D, and $80\times 80\times 80$ for the U-Net-3D+. The chosen optimizer is Adam, and after studying the optimal value for the learning rate on validation data, we set it to $10^{-3}$ in axial, coronal and 3D, and to $10^{-2}$ in sagittal. The mean absolute error (MAE) was used as the cost function, and the default value $\lambda=0.3$ for Leaky ReLU activations was retained.

The training was powered by an Intel Xeon Gold 6226R processor and a Nvidia Tesla V100 GPU.
FDK reconstruction as well as projection and back-projection operations for the iterative KL-TV algorithm, were performed using the ASTRA library \cite{vanAarle:16}. The U-Net models were implemented with Tensorflow and Keras~library. 

\subsection{U-Nets Evaluation Results}

To assess the robustness of the U-Net post-processing to outliers in the database, a Leave One Out cross-validation was applied to the entire database. In order to quantitatively compare the reconstructions, we used the Normalized Root Mean Square Error (NRMSE), the Peak Signal to Noise Ratio (PSNR) and the Structural SIMilarity (SSIM), considering the normal-dose reconstruction as ground truth. These metrics were calculated between the test volume and the reference volume, the means and standard deviations of the obtained values can be seen in Table \ref{tab:metrique_all_unet}. The U-Net-MP shows the best metrics, followed fairly closely by the axial U-Net. U-Net trained on sagittal sections performs poorly. The U-Net-3D achieves good results, slightly worse than the $\Gamma_A$ and the U-Net-MP, but with very similar values. 

The worst quantitative results in the study correspond to test volumes containing the most metal and highest HU values. In particular, dental implants and metal wires are poorly represented in our database, and generate artifacts that could not be completely removed by the proposed networks.

\begin{table*}[htp]
\centering
\resizebox{\textwidth}{!}{%
\begin{tabular}{lccccc}
\hline
\multicolumn{1}{c}{} & $\Gamma_A$              & $\Gamma_C$          & $\Gamma_S$            & U-Net MP       & U-Net 3D     \\ \hline \hline
NRMSE                & $0.0441 \pm 0.0064$ & $0.0480 \pm 0.0039$ & $0.0668 \pm 0.0483$  & $\bf 0.0432 \pm 0.0069$ & $0.0473 \pm 0.0070$ \\
PSNR                 & $56.642 \pm 1.3301$ & $55.848 \pm 1.0149$ & $53.768 \pm 3.0785$  & $\bf 56.827 \pm 1.4045$ & $56.051 \pm 1.1786$ \\
SSIM                 & $0.9982 \pm 0.0003$ & $0.9977 \pm 0.0005$ & $0.9968  \pm 0.0007$ & $\bf 0.9983\pm 0.0003$ & $0.9979 \pm 0.0003$ \\ \hline
\end{tabular}}
\caption{Means and standard deviations of evaluation metrics calculated from 32 cross-validation training sessions.} \label{tab:metrique_all_unet}
\end{table*}

\subsection{Comparison with KL-TV}

U-Nets results were compared with volumes obtained by analytical (FDK), which correspond to the input to the networks, and iterative (KL-TV) reconstruction.
We performed 500 iterations of the KL-TV algorithm, with a regularization parameter set to 0.05.

Table \ref{tab:metrics} summarizes the metrics calculated on a single volume. 
The iterative method outperforms FDK, however, deep learning allows to largely improve the FDK volumes and get the best results in our study. 

\begin{figure*}[htb]
\includegraphics[width=\textwidth]{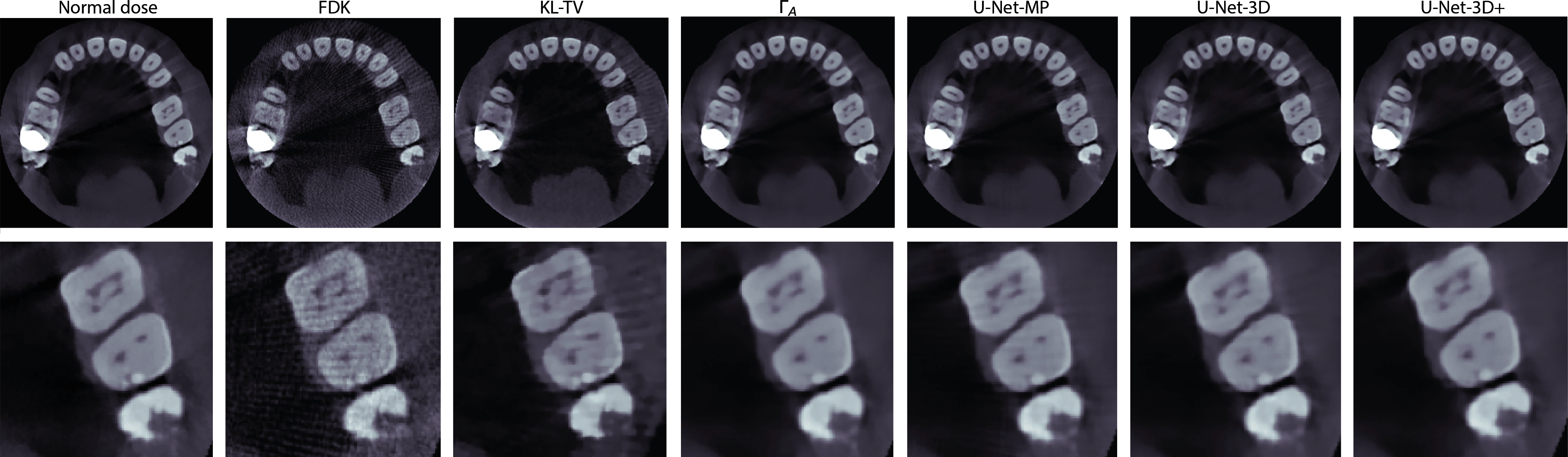}
\caption{Example of axial slice containing metal, and a zoom of the left premolar and first molar for the different methods.}
\label{fig:res}
\end{figure*}

Figure \ref{fig:res} shows an axial slice in which a crown on the left maxillary second molar generates metal artifacts. These artifacts, which are much more prevalent when few projections are used, are particularly noticeable in FDK and KL-TV reconstructions, whereas the methods based on deep learning reduce them considerably. The canals of the left first molar are better resolved with U-Net than with the other methods. 

The zoom at the center of each slice enables a better comparison of the volumes at the output of the networks. $\Gamma_A$ and U-NET-3D+ offer the best images, with effective denoising while still correctly preserving structures. Object contours in the U-Net-3D output volume are slightly less sharp than those of the others. In the volume processed by the U-Net-MP, a grid pattern is visible in the lighter areas of the teeth, due to inhomogeneities in the volume output from $\Gamma_S$. 

In terms of calculation times, it took 35 minutes to iteratively reconstruct the volumes with KL-TV. The 2D networks were trained in 1 hour and 35 minutes, the U-Net-3D in 6 hours and 18 minutes, and the U-Net-3D+ a little over 7 hours. FDK reconstruction together with 2D U-Net post-processing can be realized in less than a minute, whereas slightly less than three minutes are necessary when the 2D network is replaced with the 3D networks.


\begin{table}[!t]
\begin{center}
\begin{tabular*}{\linewidth}{@{\extracolsep{\fill}} l c c c}
\hline \hline
Method & NRMSE & PSNR & SSIM \\ \hline
FDK            & 0.169                 & 40.58                & 0.932               \\
KL-TV          & 0.136                 & 42.46               & 0.961                \\
$\Gamma_A$     & 0.037  & 53.80               & {\textbf{0.997}} \\
U-Net-MP       & 0.037  & 53.69 &  {\textbf{0.997}} \\
U-Net-3D       & 0.039            & 53.36                & 0.996                \\
U-Net-3D+ & {\textbf{0.036}} & {\textbf{53.94}}  & {\textbf{0.997}}    \\
 \hline \hline
\end{tabular*}
\end{center}
\vspace{-10pt}
\caption{NRMSE, PSNR and SSIM with the different methods on a single volume. In bold, the best value.}
\label{tab:metrics}
\end{table}

\section{Discussion and Conclusion}

Post-processing 3D volumes, reconstructed with FDK, with U-Net produces images of excellent quality, close to the ground truth. U-Net-MP and 3D fix intensity inconsistencies observed between successive slices in sagittal and coronal slices of axial U-Net reconstructions. 3D networks give the best results but lack precision when the number of parameters is too low. 

These excellent results are probably due to access to a sufficiently large database to ensure proper learning. Moreover, these methods are far less time and memory-consuming than iterative methods.

Supervised training involves learning a filter to apply to the images. As this filter is learned from the data, the composition of the database is essential. However, in our case, some features (such as metal wires) are only present in a single volume, which gives the worst in cross-validation. 

The presence of metal in some patients also complicates the normalization of volumes at network inputs. Volumes were initially normalized between -1 and 1 before passing through the networks. However, due to the high intensity of the metal, in the end, most volume values are between -0.1 and 0.1. The wide gap between voxels representing metal, and other voxels which are in the majority, can hamper training performance and diminish network results. Specific processing of the metal artifacts could be included in the reconstruction process to alleviate this problem. 

Another limitation is that we perform post-processing on FDK reconstructions, so details that are lost during reconstruction cannot be recovered. We could consider approaches that include the reconstruction process, such as method learning the reconstruction operator or unrolling methods. The first ones require very large databases and are therefore not suited to our case. The unrolling methods have already shown good results, but are tricky to implement in cone-beam geometry in terms of computational memory, even if techniques are beginning to be developed in this direction \cite{Tang2022}. Future work could investigate this line of research and the feasibility of its application in CBCT dental imaging.



\section{Compliance with ethical standards}

Ethical approval was not required as our industry partner collected and managed these data by applying the data privacy regulatory requirements.



\section{Acknowledgments}

This work was supported by the PRIMES LABEX (ANR-11-LABX-0063) at the University of Lyon, within the "Investissements d'Avenir" program (ANR-11-IDEX-0007). This work was performed using HPC resources from GENCI-IDRIS (Grant 2021-AD011012734). We would like to thank Carestream Dental for providing us with the data. 


\bibliographystyle{IEEEbib}
\bibliography{strings,refs}

\end{document}